# EmotionCarrier: A Multimodality 'Mindfulness-Training' Tool for Positive Emotional Value

Yi Wang[1, *]

1. Beijing Jiaotong University, China

## Abstract

This study introduced a Multimodal Mindfulness-Training System. Our installation, 'EmotionCarrier,' correlates traditional calligraphy interactions with real-time physiological data from an Apple Watch. We aim to enhance mindfulness training effectiveness, aiding in achieving physiological calmness through calligraphy practice. Our experiments with varied participant groups focused on data diversity, usability, and stability. We adopted methods like using EmotionCarrier for Heart Sutra transcription and adjusting installation placement for optimal user experience. Our primary finding was a correlation between calligraphy performance data and emotional responses during the transcription of the Heart Sutra.

## Introduction

With the rapid development of the large language models (LLMs), like ChatGPT, can even assist in detecting mental health patterns, enhancing the diagnosis and treatment plans[1][2]. Nevertheless, continuously exacerbating challenges such as economic recessions, escalated social conflicts, and shifting work modalities in the post-COVID-19 era all contributed to increased psychological and emotional distress among social groups[3][4]. To assist people in gaining emotional support, we have developed 'EmotionCarrier' to help people receive positive emotional feedback while relaxing their bodies and minds.

'Mindfulness' is a 'natural method' to enhance an individual's conscious awareness of their immediate sensory experiences and emotional states'[5]. Additionally, calligraphy, as a traditional art practice form, is recognized as beneficial for mental and physical relaxation[6][7][8]. Building on these foundations, we proposed an interactive installation that enables individuals to improve their 'mindfulness,' thereby obtaining positive emotional values and peaceful minds. We designed 'EmotionCarrier,' an 'interactive writing table' for individuals to express their emotions through practicing calligraphy, then incorpo- rated the 'emotional' and 'Writing' data collected into our multimodal system, providing users with tailored emotional feedback. Specifically, EmotionCarrier is a multimodality installation integrated with intelligent facilities and the large language model (Figure 1); Users' activity data from their intelligent facility and EmotionCarrier can be transmitted to ChatGPT through an API interface. Based on our developed data analysis and machine learning algorithms, we designed EmotionCarrier as a multimodal mindfulness-training tool to analyze users' emotional dynamics and provide them with visual feedback via their intelligent facilities.

In our research, we have found correlations between variations in calligraphy styles and the corresponding emotional responses. Based on our findings, this study can help to fill the gap in understanding the therapeutic benefits of traditional art forms when combined with advanced conversational AI and intelligent facilities, particularly in emotional health.

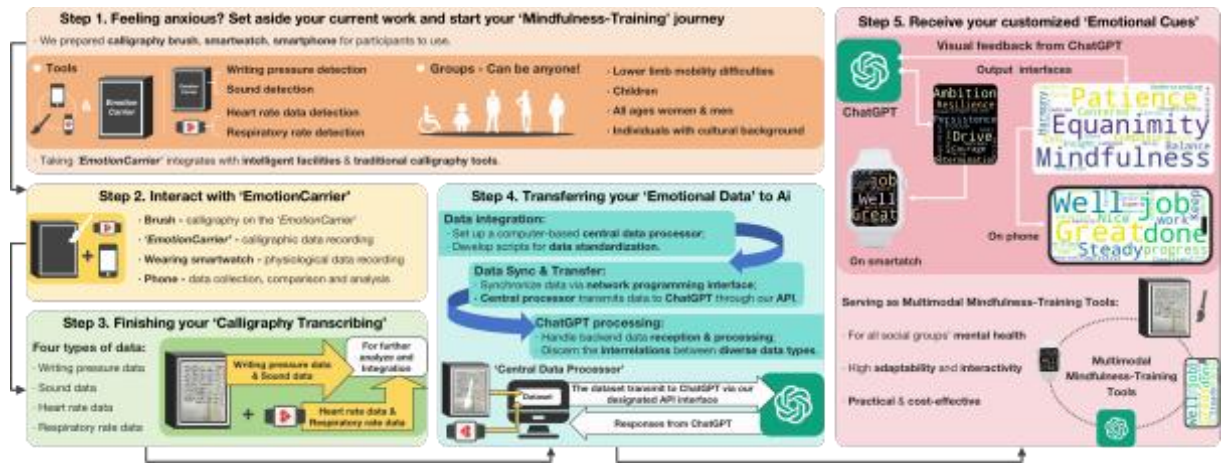

**Figure 1**: Overview of the Project's Interactive Workflow and Development Design - 'EmotionCarrier' as a Mindfulness-Training Tool.

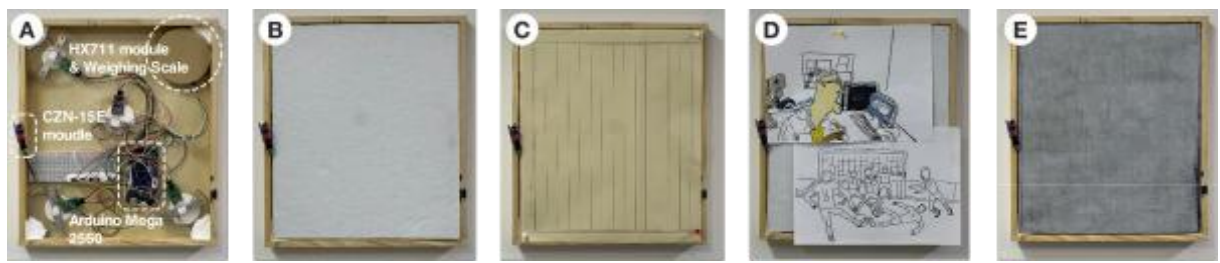

Figure 2: (A): The initial prototype of EmotionCarrier, at its core, consists of HX711 modules connected to a weighing scale, a CZN-15E module, and an Arduino Mega board. (B-E): The four different usage modes of EmotionCarrier.

## Aesthetic experience explored Emotion Carrier As A Multifunctional Interactive Medium

### 2.1. Description of 'EmotionCarrier'

'EmotionCarrier' is a multifunctional installation. Constructed with a wooden chassis, it incorporates a rigid felt board that is a versatile surface for brush writing (Figure 2-B). This board is dual-sided: one side facilitates the attachment of papers or notes with thumbtacks for calligraphy practice and work needs (Figure 2-C, Figure 2-D), while the other is a black cloth coated with a white, silica-treated eco-friendly paint (Figure 2-B). This paint becomes transparent upon wetting, mimicking the traditional ink-brush interaction on paper (Figure E); users can directly write on it with clear water, which is a more environmentally friendly way. The EmotionCarrier installation can be seen as a piece of furniture or a tool for doing everything you can imagine. We do not define the installation itself and hope that users can imagine more interesting ways to use it.

EmotionCarrier integrates three critical components for data collection in calligraphy transcribing (Figure 2-A). The HX711 modules, as bridge sensors interfaced with weighing scales, capture and output data regarding the gravitational force exerted by the calligraphy brush on the surface. The CZN-15E module, dedicated to auditory monitoring, measures the intensity of users' breathing and movement sounds.

**2.2. Revised Data Processing and Analysis with The ChatGPT**

In our research, we leveraged the powerful data processing capabilities of ChatGPT to analyze the data characteristics of users engaging in 'mindfulness-training' with EmotionCarrier. We aim to utilize

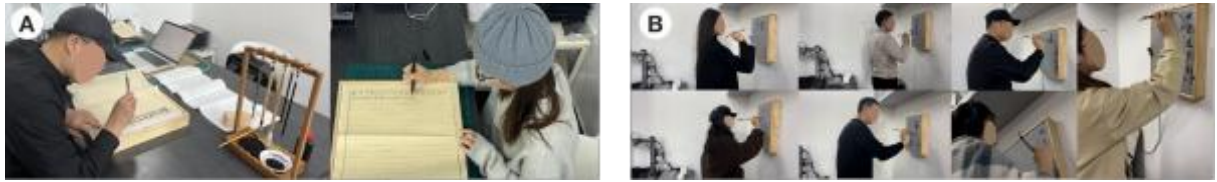

**Figure 3: (A)**: "Workshop and User Testing Experiment" - Ask each participant to operate the 'EmotionCarrier' and provide them with all related materials (e.g., smart facilities, papers, calligraphy tools, and soon.), motivating each to discover their ideal way of using it. **(B)**: "Workshop and User Testing Experiment" - Conducted user tests on the 'EmotionCarrier' installation in a workshop format, targeting different ages, occupations, and gender groups. Here, we had a total of 20 participants, including teachers (2 males, Age mean=43); designers and programmers (3 females & 3 males, Age mean=32); students (6 females & 6 males, Age mean=23).

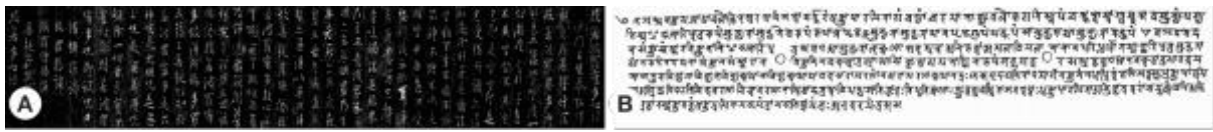

**Figure 4: (A)**: The Chinese version 'Heart Sutra', known for its exquisite calligraphy in the style of ancient Chinese calligrapher Wang Xizhi (303–361 CE). Photos via Wikipedia Commons. https://en.wikipedia.org/wik- i/Heart_Sutra **(B)**: The Sanskrit version 'Heart Sutra', believed to be the earliest extant Sanskrit manuscript (7th–8th century CE).

ChatGPT's extensive knowledge base to provide users with visual, scientific, and personalized emotional feedback.

Technically, we have designated a computer as a 'central processor,' which is responsible for receiving and managing data from EmotionCarrier and Apple Watch. The data, encompassing physiological metrics and calligraphy interaction details, is synchronized and integrated into a unified dataset via 'Alamofire', a library developed in Swift (processing language). To facilitate data communication and processing, we established a Node.js-based intermediary server that acts as an API gateway for data interconnection between our central processing system and ChatGPT's server. The key function of this intermediary server is to receive JSON-formatted data from our system, process it for compatibility, and then forward this data to the ChatGPT server for in-depth analysis (Figure 1, Step 4).

# 'Workshop and User Testing Experiment' & Data Analysis

**3.1. Design Interaction: Workshop and User Testing Experiment**

In our workshop, we engaged 20 participants from diverse professions, age groups, and genders in 'writing experiments' using EmotionCarrier for data collection and user comfort assessment. Each participant was equipped with the EmotionCarrier device and a selection of calligraphy tools, with the encouragement to experiment and determine their preferred usage method (Figure 3-A). To control experimental variables for maintaining data comparability between EmotionCarrier and Apple Watch, we offered a 'Heart Sutra' template for transcription. The Heart Sutra, renowned as one of the most eminent scriptures in Mahayana Buddhism[9], is widely utilized in practices of recitation or transcription to foster mindfulness[10], aligning well with the objectives of our study.

Considering our participants' predominantly Chinese cultural background, we provided a Chinese version of the Heart Sutra (Figure 4-A). To accommodate varied experiences in calligraphy across different languages, we also made available a Sanskrit version of the Heart Sutra for transcription purposes (Figure 4-B).

Post-experiment feedback revealed a preference for standing writing positions and using a water-dipped brush (Figure 3-B), noted for comfort, convenience, and more eco-friendliness and practicality.

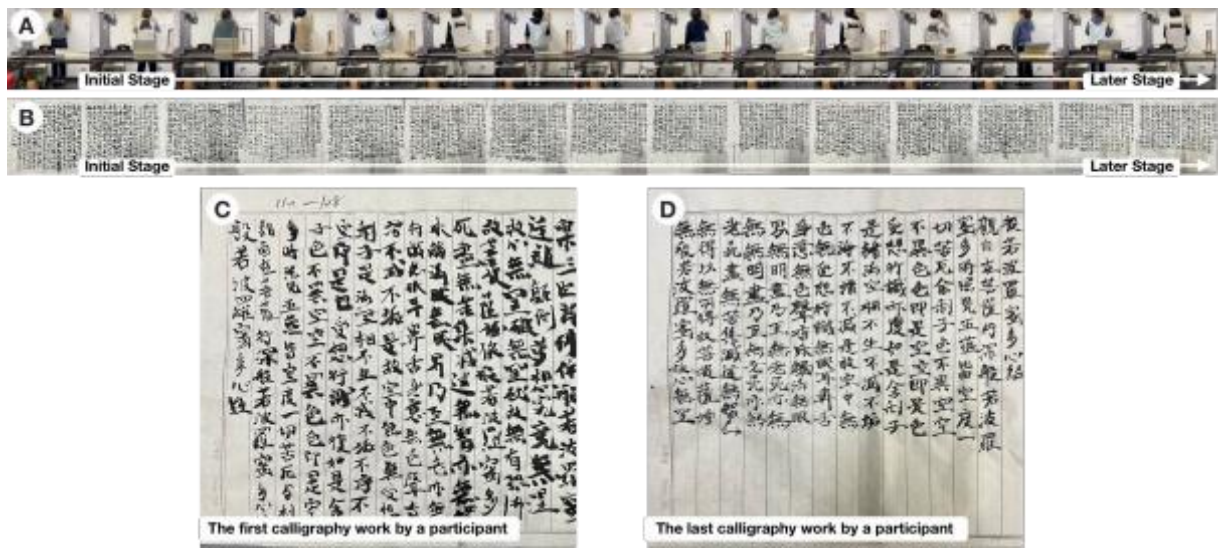

**Figure 5**: **(A)**: "Workshop and User Testing Experiment": Participants take 'EmotionCarrier' for long-lasting data collection and calligraphy practicing. **(B)**: One participant's calligraphy work, from the initial stage to the later stage. **(C)**: The first calligraphy work from one participant during the three-month experiments. **(D)**: The last calligraphy work from one participant, produced during the three-month experiments, is intended for comparison with their initial creation.

To validate the effectiveness and practicality of data collected from EmotionCarrier and Apple Watch, we undertook a series of experiments with a varied participants group, focusing on the data'sdiversity, usability, and stability during the data-collecting process. We adopted a straightforward method - fixing the paper for writing onto EmotionCarrier, and letting the user dip the brush in ink to transcribe Heart Sutra. This method allowed us to analyze the style changes of the characters that participants wrote down. Simultaneously, we adjusted the placement position EmotionCarrier based on the participants' height during the experiments, ensuring an optimal writing experience for each participant. Additionally, by adjusting the position of the CZN-15E module in EmotionCarrier closer to the user'smouth and nose, we were able to more accurately collect variations in respiratory and movement noise during the writing process.

### 3.2. Design Interaction: Data Collection & Analysis

We conducted long-term testing and data collection with participants using EmotionCarrier for writing (Figure 5-A). Over the past three months, we asked participants to wear the Apple Watch (provided by us) and use EmotionCarrier to transcribe the Heart Sutra daily, especially when they felt anxious, depressed, unsettled, or experienced other negative emotions.

Through this extensive, large-scale data collection experiment, we have our key findings:

**(1)** Participants' calligraphy work progressively improved from initial carelessness to a style with fewer mistakes, characterized by neatness and standardization in the past three months (Figure 5-B, C&D).

**(2)** Physiological data from the Apple Watch indicated a trend towards stabilization during the Heart Sutra transcription.

**(3)** Analysis of 30 varied samples from different experiment stages revealed:

• A negative correlation between writing pressure and duration in Heart Sutra transcriptions using EmotionCarrier (Figure 6-A). This means that the longer the duration of continuous transcribing by the participants, the greater and more stable the writing pressure becomes, indirectly reflecting a gradual stabilization of the participants' state throughout the writing process and viceversa.

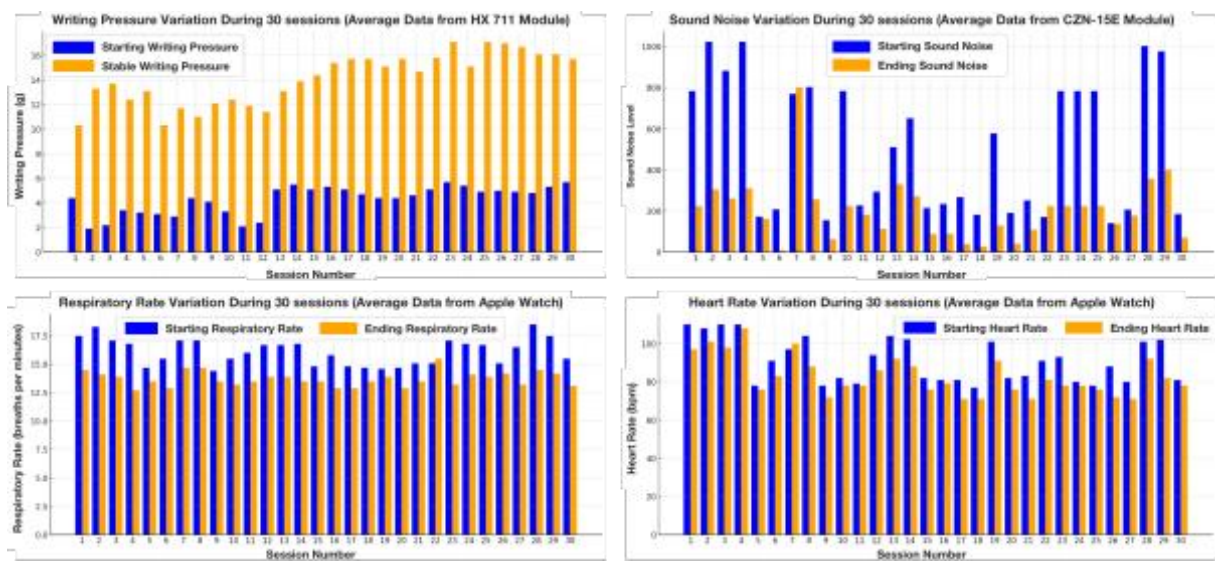

**Figure 6**: **(A)**: Writing Pressure Variation During 30 Sessions (Average Data from HX 711 Module; **(B)**: Breath Sound & Body Movement Noise During 30 Sessions (Data from CZN-15E Module); **(C)**: Respiratory Rate Variation During 30 Writing Sessions (Data from Apple Watch); **(D)**: Heart Rate Variation During 30 Sessions (Data from Apple Watch).

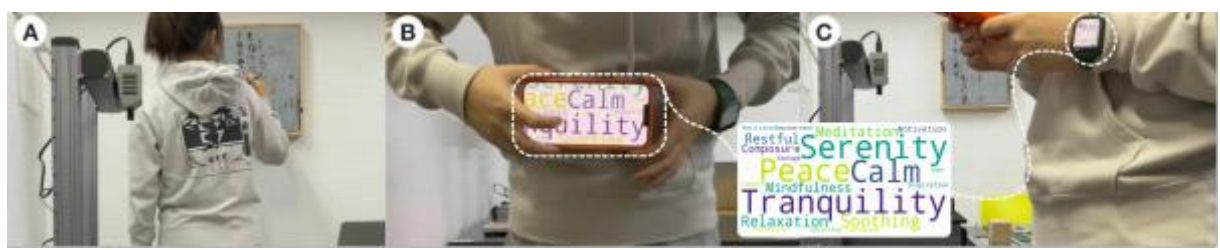

**Figure 7: (A):** Participant use EmotionCarrier for transcribing Heart Sutra. **(B&C)**: After finishing practicing calligraphy for a while, the participant received the 'Emotion Feedback' from the intelligent communication devices - Apple Watch and Phone.

• A similar negative correlation was observed in respiratory and movement noise during these transcriptions (Figure 6-B). This implies that the longer the participants continue to transcribe, the lower and more stable their respiratory and movement noise becomes, indirectly indicating a gradual stabilization of the participants' state throughout the writing process and viceversa.

• A trend of decreasing and stabilizing respiratory rates post-transcription of the Heart Sutra by using EmotionCarrier (Figure 6-C).

• A comparable trend in heart rate stabilization post-transcription (Figure 6-D). The variation in the heart and respiratory rate data suggests that the overall physiological state of participants stabilizes during the transcription process.

## Interactive Experience & Future Challenges

Through participant involvement in our experiments, we demonstrated the effectiveness of calligraphy practice in mindfulness training, which aids in achieving physiological calmness. We progressed to integrate the functionalities of EmotionCarrier, Apple Watch, and ChatGPT in further experiments. This integration was undertaken with the anticipation that the extensive knowledge base and analytical capabilities of large language models like ChatGPT would offer users valuable emotional visual feed- back, thereby enhancing the overall interaction experience in the mindfulness-training process using EmotionCarrier.

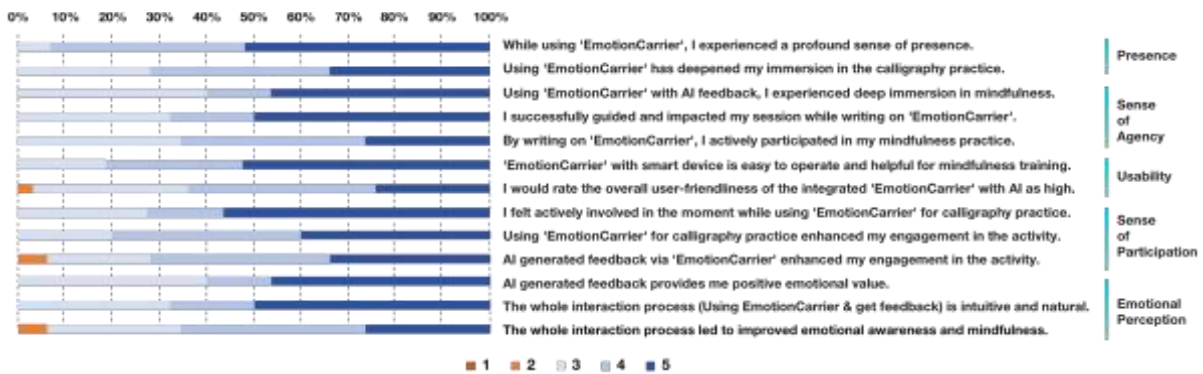

**Figure 8:** Evaluation from the 20 participants who involved in our experiments on 'interacting with the Emotion- Carrier'. The rating shows participants on 'Presence', 'Sense of Agency', 'Usability', 'Sense of Participation', and 'Emotional Perception'. (1=Strongly disagree, 5=Strongly agree).

In our subsequent experiments, we transferred interaction data from EmotionCarrier to the Chat- GPT server via an established API gateway. This data included detailed descriptions of each dataset component, such as the interrelations between physiological metrics (heart rate and respiratory rate) and writing duration, along with the models of our intelligent devices. Utilizing this data, ChatGPT developed a user interface (UI) for our intelligent facilities (Apple Watch and smartphone), equipped with 'emotional prompt words' for post-interaction visualization with EmotionCarrier (Figure 7-A, B&C). We shared these interfaces with our participants through their intelligent facilities, receiving affirmative feedback from them. Our participants reported that this 'multimodal' interaction experience was engaging and contributed to enhanced relaxation, introspection, and mindfulness (Figure 8). Our study, by employing common Apple products and the large language model ChatGPT, effectively demonstrated the success of our multimodal interaction system (including EmotionCarrier and intelligent facilities) in facilitating mindfulness training.

## Future research

Future studies can further investigate a number of interesting research directions described in these papers. It is well known already that personalized and adaptive systems, such as context-aware wearables and emotionally intelligent interfaces, can enhance the user experience by adapting to individual needs in real-time. Inclusive and accessible principles can be used in designing interactive products and systems for users with different abilities and cultural backgrounds. Multisensory input and biofeedback mechanisms can help to improve user engagement and emotional well-being. Co-design and user participation methods can ensure that user values and preferences are considered sufficiently in the design process, fostering empathy-driven design.

When we called for papers with the word "connectivity" in the title, we expected the submissions to address more the distributed systems that would involve interconnected products and systems, and further people and their societies. However, almost none of the papers addressed such a topic. There are several aspects were not well-covered in these papers. None of the papers undertook longitudinal studies that are necessary to evaluate the long-term impact of connectivity and related aesthetics and technologies on behavior, emotional health, and social dynamics. Most of these papers did not discuss ethical and sustainable issues, such as environmentally friendly materials and ethical AI practices. These issues should be addressed more in the design and research processes.

These research directions will contribute to designing more inclusive, adaptive, and user-centered products, systems, and services by addressing the evolving needs and challenges in the fields of connectivity, aesthetics, and empowerment.